\documentclass[aps,prb,reprint,groupedaddress]{revtex4-1}

\pdfoutput=1

\usepackage{graphicx}
\usepackage{dcolumn}
\usepackage{bm}
\usepackage{braket}
\usepackage{float}
\usepackage{amsmath}

\begin{document}

\newcommand{\smb}{SmB$_{6}$}
\newcommand{\kt}{k_B T}

\title{Understanding low-temperature bulk transport in samarium hexaboride without relying on in-gap bulk states}

\author{A. Rakoski}
\email[]{ralexa@umich.edu}
\affiliation{University of Michigan}

\author{Y.S. Eo}
\affiliation{University of Michigan}

\author{K. Sun}
\affiliation{University of Michigan}

\author{\c{C}. Kurdak}
\affiliation{University of Michigan}

\date{\today}

\begin{abstract}
We present a new model to explain the difference between the transport and spectroscopy gaps in samarium hexaboride (\smb), which has been a mystery for some time. We propose that \smb{} can be modeled as an intrinsic semiconductor with a depletion length that diverges at cryogenic temperatures. In this model, we find a self-consistent solution to Poisson's equation in the bulk, with boundary conditions based on Fermi energy pinning due to surface charges. The solution yields band bending in the bulk; this explains the difference between the two gaps because spectroscopic methods measure the gap near the surface, while transport measures the average over the bulk. We also connect the model to transport parameters, including the Hall coefficient and thermopower, using semiclassical transport theory. The divergence of the depletion length additionally explains the 10-12 K feature in data for these parameters, demonstrating a crossover from bulk dominated transport above this temperature to surface-dominated transport below this temperature. We find good agreement between our model and a collection of transport data from 4-40 K. This model can also be generalized to materials with similar band structure. 
\end{abstract}

\maketitle

\section{Introduction}
Samarium hexaboride (SmB$_6$) has long eluded classification due to its unique properties, beginning with its unusual temperature-resistivity curve.\cite{menth} Past research has led to the classification of \smb{} as a rare-earth mixed valence compound; although it is homogeneous, the Sm ions have mixed valence of 2+ and 3+.\cite{menth, cohen, maple} \smb{} has a simple cubic structure, and it is also free of both magnetic and structural phase transitions, \cite{menth} which makes it a good candidate for studying homogenous mixed valence. One of the most significant developments was the discovery that \smb{} is a Kondo insulator. \cite{aeppli} Kondo insulators are characterized by the opening of a small gap at the Fermi energy due to hybridization between $f$-electrons and conduction electrons. \cite{kondo, coleman} In \smb{} specifically, there are three $4f$ bands, one of which hybridizes with the $5d$ conduction electrons, opening a gap between the two hybrid bands. \cite{mott}

Although the presence of the hybridization gap suggests that \smb{} should exhibit insulating behavior, detailed transport results over a 40-year period suggest a much more complicated picture. \cite{menth, nickerson,allen,molnar,cooley,wolgast} High-quality \smb{} has consistently demonstrated activated behavior from 4-40 K. Around 4 K, the activated behavior is always terminated by an unknown conduction mechanism, leading to a plateau in the temperature-resistivity curve. Many attempts have been made to explain this crossover to conductive behavior in terms of bulk effects, with a commonly accepted picture based on impurities in the material.\cite{menth} A breakthrough came from the prediction that Kondo insulators can additionally be topological insulators, \cite{dzero, takimoto, dzerob} a class of materials that undergo a crossover at low temperature from a conventional state to a bulk insulating state with topologically protected metallic surface states. \cite{fukane} Indeed, recent transport \cite{wolgast, kim} and angle-resolved photoemission spectroscopy (ARPES) results \cite{denlinger,neupane,xu} have demonstrated evidence for these topologically protected metallic surface states in \smb{} with a crossover temperature of approximately 4 K. In this paper, we will concentrate on temperatures from 4-40 K, where bulk transport is dominant. Although activated behavior is consistently observed in this temperature range, transport measurements including resistivity and Hall coefficient also demonstrate a feature at about 10 K.

Before discussing some of the unusual properties of \smb{} that arise from its small gap and hybridized bands, it is instructive to discuss bulk transport in standard TIs.  Such ``standard'' TIs (for example, Bi$_2$Se$_3$ or Bi$_2$Te$_3$) are characterized by a bulk band gap as in conventional semiconductors, and after undergoing the crossover to the topological phase, there are an odd number of surface states located in the gap. \cite{hasan, qizhang} If the topological surface states are not considered, all standard TIs can be treated as semiconductors. In this picture, charge neutrality must be enforced, so the Fermi energy ($E_F$) is initially expected to be exactly halfway between the top of the valence band and the bottom of the conduction band. As in semiconductors, impurity states may also be present in the gap. For an $n$-type material, donor states would be in the gap near the conduction band, and for a $p$-type material, acceptor states would be in the gap near the valence band. Because charge neutrality must also be enforced, the presence of these extra states shifts $E_F$ towards the conduction band for donor states and towards the valence band for acceptor states. \cite{sze}

Impurity states in semiconductors and standard TIs can be treated quantitatively using the effective mass approximation. \cite{kohn} In this picture, impurities are assumed to be hydrogenic, but with the substitution of effective mass for electron mass ($m \rightarrow m^{*}$) and dielectric constant for vacuum permittivity ($\varepsilon_0\rightarrow \kappa \varepsilon_0$). The results are an effective Bohr radius
\begin{equation}
\label{bohr}
a_B^{*} = \frac{4 \pi \kappa \varepsilon_0 \hbar^2}{m^{*} e^2} = \frac{\kappa}{m^{*}/m} (0.53 \: \textrm{\r{A}})
\end{equation}
and an effective ground-state energy 
\begin{equation}
\label{gsen}
E^{*} = - \frac{m^{*} e^4}{2 (4 \pi \kappa \varepsilon_0)^2 \hbar^2} = - \frac{m^{*}/m}{\kappa^2} (13.6 \: \textrm{eV}).
\end{equation}
Once these parameters have been calculated, the initial assumption of hydrogenic impurities can be verified. In a donor state, for example, the extra electron must have an extent much larger than one lattice constant for the donor to be hydrogenic.  If this were not the case, contributions from the donor itself would also have to be considered, and the hydrogen model could not be used. Therefore, when comparing the effective Bohr radius to the lattice constant, we must satisfy the condition $a_B^{*} \gg a$ for the approximation to be valid. 

The effective mass approximation has been used successfully in all standard TIs. Standard TIs exhibit residual bulk conduction after undergoing the crossover to the topological state, and this is well-understood as arising from impurity states that can be treated using the effective mass approximation. Many researchers have applied the idea of impurity states in the gap to \smb{} to try to understand the plateau at 4 K before it was thought to be a TI, \cite{menth} and later to explain experimental discrepancies in the size of the Kondo hybridization gap.  While the presence of the gap is well-known, transport and spectroscopic methods disagree on the size. Transport measurements, which probe the energy difference between $E_F$ and the conduction band (the activation energy), report 3-4 meV. \cite{allen, molnar,cooley, kim, wolgast} In analogy with semiconductors and standard TIs, it is expected that $E_F$ is exactly halfway between the valence and conduction bands, suggesting that the total transport gap is 6-8 meV. On the other hand, spectroscopy and tunneling experiments measure the full gap near the surface, and they report 16-20 meV.\cite{flachbart, gorshunov, jiang, ohta, nanba, travaglini, zhang} 

This discrepancy has been interpreted as arising from the presence of in-gap bulk impurity states.\cite{mott, travaglini, ohta, nanba, sluchanko99, gorshunov, dressel, flachbart, rossler} In this scenario, transport would measure the difference between the impurity state and the conduction band,yielding an incorrect result for the total gap. Since the effective mass approximation is usually so successful, it has been applied to \smb{} to understand this proposed in-gap impurity state. However, reported values \cite{gorshunov, travaglini} of the dielectric constant $\kappa$ range from 600-1500, and as we have seen, activation energy (which can be used to obtain effective mass) ranges from 3-10 meV. Using these values with Eqs. \ref{bohr} and \ref{gsen}, we obtain a minimum $a_B^{*}$ of 0.5 \r{A} and a maximum $a_B^{*}$ of 4 \r{A}. Reports of the effective Bohr radius are usually in this range; for example, Sluchanko $et \: al.$ report 3 \r{A}.\cite{sluchanko99} Additionally, most reports agree that the lattice constant $a$ of \smb{} is about 4.13 \r{A}, so we find that the condition $a_B^{*} \gg a$ required to verify the effective mass approximation is not satisfied anywhere in the range of Bohr radii that can be calculated. Because the effective mass approximation fails, in-gap impurity states in \smb{} are not justified. 

Since semiconductor theory and the effective mass approximation are successful in most cases, this result is startling. However, upon closer examination, we find that it is perhaps not completely unexpected. Because the gap in \smb{} arises due to hybridization, its band structure is very different from that of a conventional semiconductor. Unlike a semiconductor, \smb{} has a non-quadratic and asymmetric dispersion, because its band structure arises due to Kondo hybridization. Both the valence and conduction bands have nearly flat regions characterized by the localized $f$ states as well as low-mass regions characterized by the $d$ states. Because of this unusual composition, while the gap and band structure effects in \smb{} arise based on contributions from all the carriers, transport is dominated only by the low-mass carriers. Additionally, both bands have positive curvature, unlike in a semiconductor, where only the conduction band has positive curvature. As we will see, this has a significant effect on how we understand transport. The gap is also much smaller than that of standard semiconductors or TIs. Because of these differences in the band structure, we will see that \smb{} must be treated much more carefully than standard gapped materials. (In contrast, other hexaboride materials can be treated as standard gapped materials, and in these cases, impurity states are present within the bulk gap. \cite{stankiewicz})

In the context of in-gap bulk states, we can gain insight into \smb{} by analogy with superconductors. When a material undergoes a transition to a superconducting state, some of the electrons near the Fermi energy condense into Cooper pairs. Formation of the condensate opens up a gap at the Fermi energy. \cite{bcs} Even though this gap is so small, tunneling measurements \cite{reif} have shown that the addition of impurities to a superconductor does not destroy superconductivity (until the impurity concentration becomes sufficiently high). This suggests that the impurity states are not in the gap, or that superconductors exhibit a small and clean gap and are not vulnerable to impurity conduction. Although the mechanism for gap formation (the Kondo effect) is very different in \smb{}, the gap is also much smaller than a semiconductor bulk gap. This, combined with the failure of the effective mass approximation, suggests that \smb{} should have a clean gap that is not vulnerable to impurity conduction at low temperatures. In fact, \smb{} does not exhibit residual bulk conduction experimentally, and this can be taken as evidence for a clean bulk gap. \cite{wolgast}

Another interpretation of the gap discrepancy is that spectroscopy measures the direct gap while transport measures an indirect gap that forms during hybridization. Theoretical treatments of Kondo hybridization predict the presence of both a direct and an indirect gap, \cite{coleman, alexandrov, lu} and researchers have also used this idea to explain the gap discrepancy. \cite{cooley, zhang} This interpretation explains the gap discrepancy while avoiding the problem of the in-gap impurity states, but we will propose an alternative explanation that is consistent with features observed in transport. As we have seen, in-gap impurity states in \smb{} are not justified by the effective mass approximation, nor are they consistent with the observation of no residual bulk conduction after the surface states become relevant. To achieve consistency, we propose another explanation of the gap discrepancy that does not rely on in-gap states and instead allows for a clean direct gap.

\section{Simplified density of states and band bending calculations}
\subsection{Density of states}
\begin{figure}
\includegraphics[scale=1, trim = 2mm 4mm 0mm 0mm, clip]{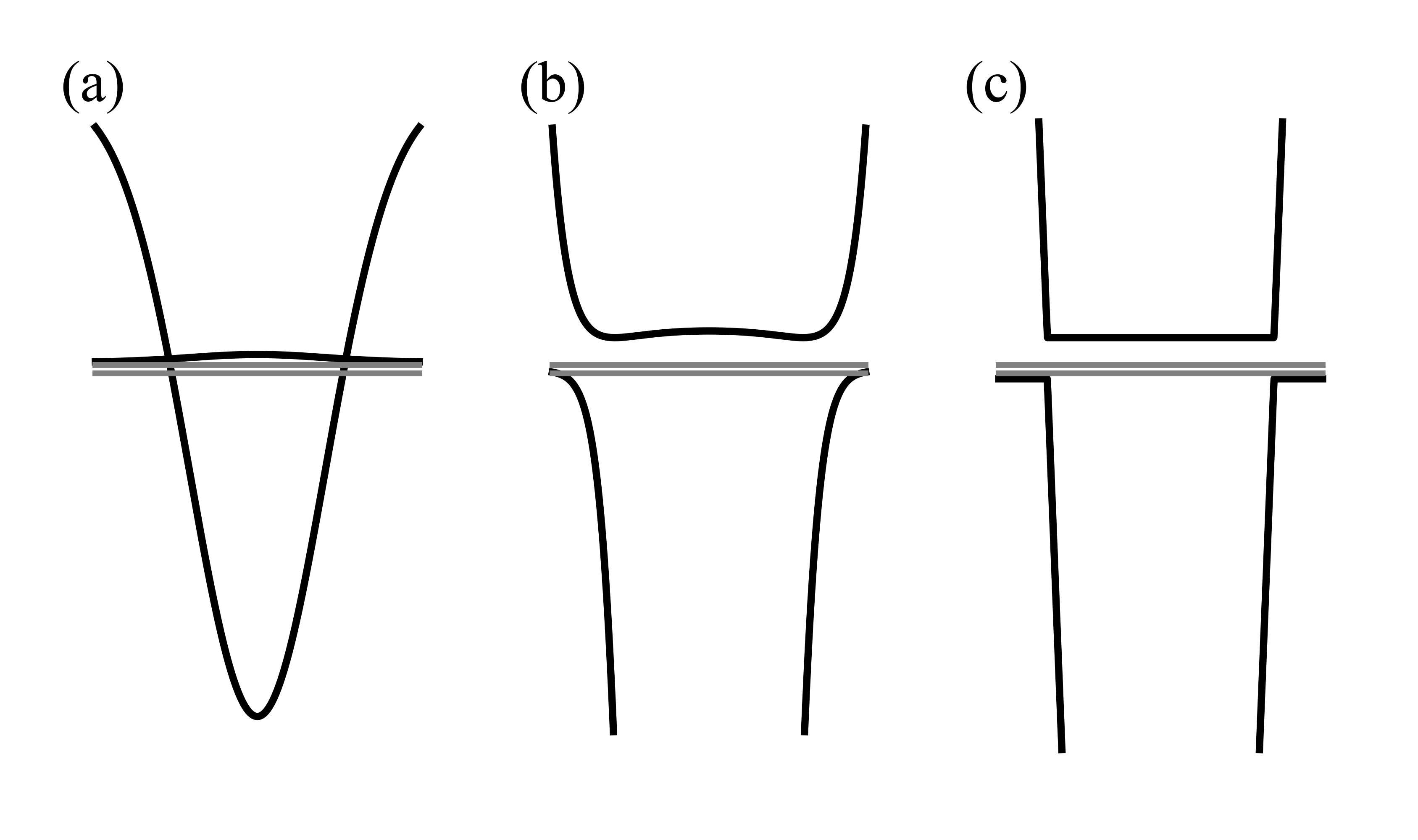}%
\caption{ Dispersion relation of \smb{} along the $\Gamma - X - \Gamma$ direction, focused where the hybridization takes place. (a) Band structure before hybridization. (b) Band structure after hybridization of the $d$ band with one of the $f$ bands. The vertical axis (energy) scale is zoomed in from the scale in (a). (c). Simplified band structure used in the calculation. The vertical scale is the same as in (b).  \label{fig1}}
\end{figure}

The dispersion in \smb{} is well-known based on the success of recent high-resolution ARPES measurements.\cite{denlinger,neupane,xu} The gap forms at low temperature, when the conduction band (5$d$) hybridizes with localized states (4$f$). In \smb{}, there are three 4$f$ bands, and ARPES indicates that only one band participates in hybridization. A sketch of the band structure of \smb{} before hybridization is shown in Fig. \ref{fig1}(a), and the hybridized band structure is shown in Fig. \ref{fig1}(b), both along the $\Gamma - X - \Gamma$ direction. \smb{} is an insulator, so the Fermi energy is located in the gap in the hybridized band structure. As can be seen in the figure, the two unhybridized $f$ bands are still present. 

In addition to this basic structure, the hybridized dispersion has some subtle features, which can also be observed by ARPES. For example, ARPES demonstrates an indirect gap with a valence peak about 15 meV below the Fermi energy at the $H$ point. \cite{denlinger} This feature can also be observed at nearby energies due to intrinsic and thermal broadening. Compared to the main features of the dispersion, however, this feature is small and close to the valence band. Although ARPES cannot probe far into the conduction band at the temperatures we are considering, there are likely some similarly small features present in the conduction band. We refer to the regions in which such small features exist as the ``region of non-parabolicity.''

In our model, we will use a dispersion that is simplified considerably from the actual dispersion. We neglect the small features in the region of non-parabolicity, such as the feature at the $H$-point and any similar features in the conduction band. To do this, we approximate the band structure using a piecewise function, as shown in Fig. \ref{fig1}(c). Here, the flat regions approximate the pieces of the hybridized dispersion that primarily come from the $4f$ band, which we refer to as ``$f$-like'' states. The linear regions approximate the pieces of the hybridized dispersion that primarily come from the $5d$ band, which we call ``$d$-like'' states. Additionally, the two unhybridized $4f$ bands cannot be resolved separately from the valence band by ARPES, so we approximate them to be at the top of the valence band. Making these approximations introduces some error into the model, but the features in the region of non-parabolicity are small, so the error is not more than a few meV. 

\begin{figure}
\includegraphics[scale=1, trim = 2mm 4mm 0mm 0mm, clip]{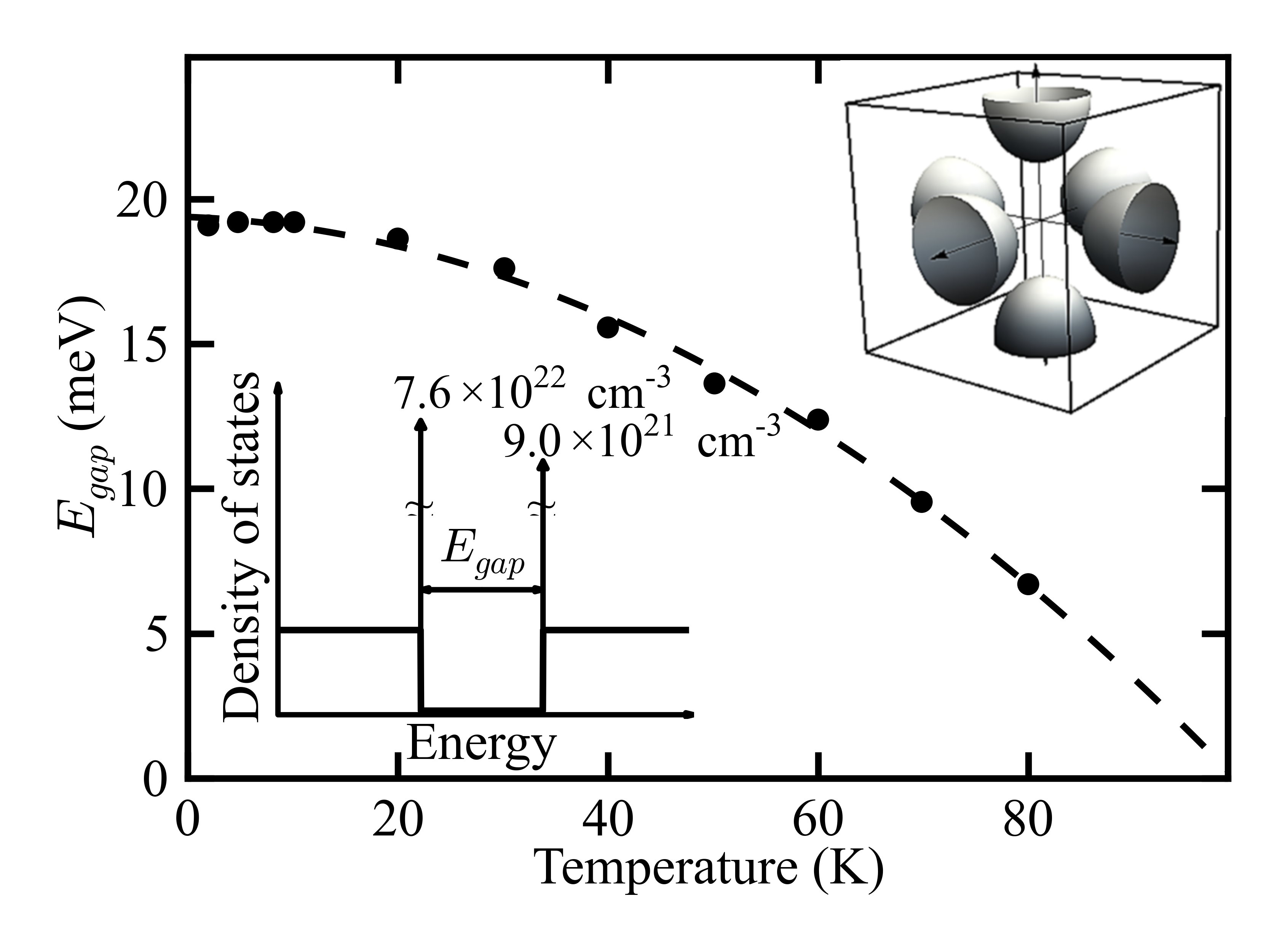}%
\caption{Parameters used in the calculation. Main plot: Data \cite{zhang} for dependence of the gap size on temperature and a best fit (dashed line). Lower left inset: Simplified density of states used in the calculation. Upper right inset: Fermi surface and Brillouin zone of \smb{} after hybridization. \cite{pastandpresent} \label{fig2}}
\end{figure}

From this dispersion, we can calculate a simplified density of states (DOS). The DOS corresponding to our simplified dispersion (Fig. \ref{fig1}(c)) is shown in the lower inset of Fig. \ref{fig2}. In this figure, the peaks in the DOS correspond to the $f$-like regions of the dispersion, and these can be estimated from the size of the pockets in the \smb{} Brillouin zone (BZ), shown in the upper inset of  Fig. 2. The flat parts of the DOS correspond to the $d$-like regions of the dispersion. In the range $E_{gap}$, the DOS is zero, and the gap changes with temperature. Data for the gap as a function of temperature \cite{zhang} is shown in Fig. 2, as is the fit to this data that was used in the calculation (dashed line).  

We can represent the simplified DOS using delta functions ($\delta$) for the $f$-like states and step functions ($\theta$) for the $d$-like states:
\begin{equation}
\begin{split}
\label{dos}
g(\epsilon) & = N_{cf} \delta(\epsilon - E_C) + g_{cd} \theta(\epsilon - E_C) \\
&+N_{vf} \delta(\epsilon - E_V) + g_{vd} (\theta(\epsilon - E_V) + 1)
\end{split}
\end{equation}
where $E_C$ is the edge of the conduction band, $E_V$ is the edge of the valence band, $N$ is a density in cm$^{-3}$, $g$ is a DOS in cm$^{-3} \cdot$ eV$^{-1}$, and the Fermi energy $E_F$ has been set to 0. The subscripts on the four factors refer to electron type and band; e.g. $N_{cf}$ is the density of states for $f$-like electrons in the conduction band ($c$). The Fermi energy, which is in the gap, is defined to be zero. 

The sizes of these four terms can be estimated using ARPES data. \cite{pastandpresent} According to this measurement, the hybridized $f$ band BZ has six half-ellipsoid Fermi pockets (upper inset of Fig. \ref{fig2}), and the total volume of these yields the number of filled states in the conduction band, which we will denote $n_{ell}$. This implies that 
\begin{equation}
\label{ncf}
N_{cf} = n_{ell} = 9.0 \times 10^{21}  \: \textrm{cm}^{-3}.
\end{equation}
The remaining volume in that BZ, plus the total volume of the BZs for the two unhybridized $f$ bands, yields the valence band contribution. We will denote the volume of the BZ as $n_{BZ}$, so this implies that
\begin{equation}
\label{nvf}
N_{vf}= 3 n_{BZ} - n_{ell} =7.6 \times 10^{22} \: \textrm{cm}^{-3}.
\end{equation}
The $d$-like states can be calculated from data above the hybridization temperature.\cite{pastandpresent} We approximate the dispersion to be quadratic and use the usual result for a 3D quadratic DOS, 
\begin{equation}
\label{dos3d}
g_{3D}(\epsilon) = \frac{m}{\pi^2 \hbar^3} \sqrt{2 m \epsilon}
\end{equation}
In our simplified DOS we can approximate the $d$ bands on both sides of the gap as constant. Specifically, we approximate this constant to be $g_{3D}(E_{F}$), because $E_F$ is in the gap and the gap is small. This also means that this value in both bands is about the same constant, 
\begin{equation}
\label{ncd}
g_{cd} \approx g_{vd} \approx  g_{3D}(E_F) = 1.8 \times 10^{19}  \: \textrm{cm}^{-3} \: \textrm{eV}^{-1}.
\end{equation}
We note that this term is 2-3 orders of magnitude smaller than the terms in Eqs. \ref{ncf} and \ref{nvf}.

\subsection{Band structure calculation}
As we have seen, the actual and simplified dispersions as well as the DOS of \smb{} can be characterized by two types of carriers. The flat regions are dominated by $f$-like carriers, and the remainder is dominated by $d$-like carriers. In this section, we will outline a self-consistent calculation used to obtain the band structure. For such a calculation, we must take all the charges into account. However, the $f$-like terms (Eqs. \ref{ncf} and \ref{nvf}) are 2-3 orders of magnitude greater than the $d$-like coefficients (Eq. \ref{ncd}). To get the total charge density, the $d$-like states require a factor of $k_B T$, so they become even smaller; because of this we neglect the $d$-like carriers for the band structure calculation. However, we will later see that transport is governed by the low-mass, $d$-like carriers.

The charge density can be calculated using usual methods for semiconductors. In semiconductors, the conduction band is nearly empty, so the Fermi-Dirac distribution $f^0(\epsilon)$ can be approximated by the Boltzmann distribution. The electron density is 
\begin{equation}
\label{densn}
n = \int_{E_C}^{\infty} f^0 (\epsilon) g(\epsilon) d\epsilon = n_0 \: e^{-(E_C - E_F)/\kt}
\end{equation}
where $n_0$ is the average DOS. In \smb{} we use Eq. \ref{dos} for the DOS, keeping only the delta function terms. This yields electrons in the conduction band with approximate density
\begin{equation}
n \approx N_{cf}  \: e^{-(E_C - E_F)/k_B T}.
\end{equation}
We can similarly calculate the approximate density of holes in the valence band to be
\begin{equation}
p \approx N_{vf} \: e^{-(E_F - E_V)/\kt}
\end{equation}
where $E_V$ is the valence band edge. This result resembles the carrier density of a conventional semiconductor. For such semiconductors, charge neutrality, $n = p$, yields the intrinsic carrier density
\begin{equation}
n = p = n_i \approx \sqrt{N_{cf} N_{vf}} \: e^{-E_{gap}/2\kt}.
\label{intrinsic}
\end{equation}

The intrinsic picture works well for \smb{} at high temperatures. However, at sufficiently low temperature, the TI surface states become relevant, and the related surface charges start to contribute. Requiring charge neutrality forces the Fermi energy to be pinned in place, leading to band bending in the valence and conduction bands. This possibility has been suggested by recent experimental results, \cite{ishida} but was not previously explored in depth.

To understand the effects of band bending, we perform a self-consistent calculation to obtain the band structure. In this calculation, we model the effects of band bending using a potential $\phi(z)$ of the form
\begin{equation}
e \phi(z) = E_C (z) - E_{gap} /2
\label{pot}
\end{equation}
where the conduction band is now dependent on location $z$ in the bulk, and $E_{gap}=E_C-E_V$. We can rewrite the carrier densities in terms of this potential to obtain
\begin{equation}
n(z) =N_{cf} \exp{\bigg[- \frac{e \phi (z) +  E_{gap}/2 - E_F}{\kt} \bigg]} 
\end{equation}
and a similar expression for $p(z)$. Using charge neutrality again, we obtain 
\begin{equation}
\rho (z) = - e n(z) + e p (z) = 2 n_i e \sinh{\bigg[ \frac{e \phi (z)}{\kt} \bigg] }
\label{rhoz}
\end{equation}
for the total charge density. We then solve for the potential across the bulk using Poisson's equation in one dimension,
\begin{equation}
\label{poissonsolve}
\frac{d^2 \phi}{dz^2} = \frac{ \rho (z)}{\varepsilon} = \frac{2 n_i e}{\varepsilon} \sinh{\bigg[ \frac{e \phi (z)}{\kt} \bigg]}.
\end{equation}

To solve this equation, we choose a ``test sample'' of thickness 200 $\mu$m, which is typical to a real \smb{} sample. We define $z=0$ as the center of the sample (so that $z= \pm 100 \: \mu$m are the edges). In addition, we use $\kappa = 600$ as the dielectric constant. \cite{gorshunov} To proceed with the solution, we now require boundary conditions. For the first boundary condition, we simulate the band bending effects on the surface by introducing a pinning energy $E_{pin}$, which is defined with respect to the midpoint of the gap by $E_{pin} = (E_C + E_V)/2 - E_F$. This describes the energy difference between the band at zero potential and the minimum of the bent band. Therefore, the boundary condition can be expressed as $e \phi(z =100 \: \mu m) = - E_{pin}$. We can also define this pinning relative to the gap as $E_{pin} = E_{gap}/2 - E_a$. As we will see later, when temperature is sufficiently low, $E_a$ corresponds to the activation gap measured by transport. The pinning energy must also be the same at both edges of the sample, and to enforce this, the second boundary condition is that $d \phi / dz = 0$ at $z=0$, the center of the sample. 

\begin{figure}
\includegraphics[scale=1, trim = 2mm 4mm 0mm 0mm, clip]{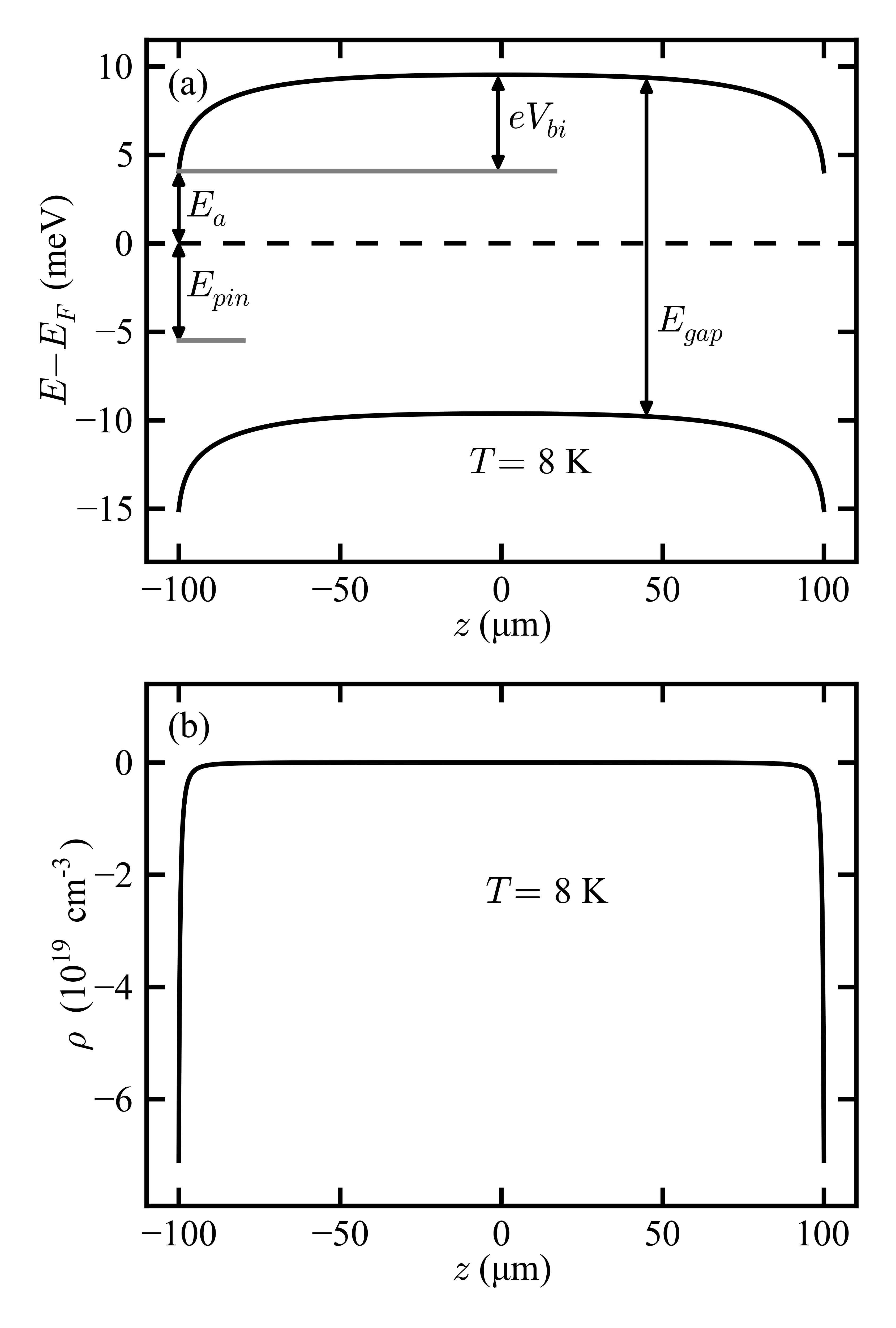}%
\caption{Parameters obtained from the self-consistent solution for $\phi(z)$ in a 200 $\mu$m sample with $E_{pin} = 5.5$ meV and $T = 8$ K. (a) Band structure obtained using Eq. \ref{pot}. The relationship among the activation energy $E_a$, the built-in potential $eV_{bi}$, the pinning $E_{pin}$, and the gap $E_{gap}$ are all shown. (b) Charge density obtained using Eq. \ref{rhoz}.
 \label{fig3}}
\end{figure}

Additionally, we can define a ``built-in potential,'' $V_{bi}$, as is commonly done for band-bending calculations in semiconductors. \cite{sze} $V_{bi}$ describes the magnitude of the bending in terms of the difference between the maximum and minimum points on the conduction or valence band. Using this built-in potential, we can determine an associated length scale (the depletion length) of Eq. \ref{poissonsolve}, given by
\begin{equation}
l = \sqrt{\frac{2 \epsilon V_{bi}}{e n_i}} = \sqrt{\frac{2 \epsilon V_{bi}}{e (N_{cf} N_{vf})^{1/2}}} \: e^{E_{gap}/4\kt}.
\label{lengthscale}
\end{equation}
At low temperature, when $E_{gap}$ is large, the depletion length is large, and at high temperature, when $E_{gap}$ is small, the depletion length is small. 

With these parameters, solutions to Eq. \ref{poissonsolve} were found for temperatures of 4-40 K and values of $E_{pin}$ between 4 and 7 meV. From a solution $\phi(z)$, the conduction and valence bands can be obtained from Eq. \ref{pot}, and the charge density can be obtained from Eq. \ref{rhoz}. An example of these are shown for 8 K and $E_{pin} = 5.5$ meV in Fig \ref{fig3}. Fig. \ref{fig3}(a) shows the calculated conduction and valence bands, as well as the relationships among the band structure and the parameters $E_{pin}$, $E_{gap}$, $E_a$, and $eV_{bi}$. We note that the valence band is always parallel to the conduction band and can be obtained by subtracting $E_C(z)-E_{gap}$. Because of this symmetry, the following discussion will be confined to the conduction band, although it will also apply to the valence band. Fig. \ref{fig3}(b) shows the calculated charge density corresponding to this band structure. Across the sample, the charge density is negative, and its magnitude is largest near the surfaces. This is expected, because excess charge at the surfaces leads to band bending. 
 
\begin{figure}
\includegraphics[scale=1, trim = 2mm 4mm 0mm 0mm, clip]{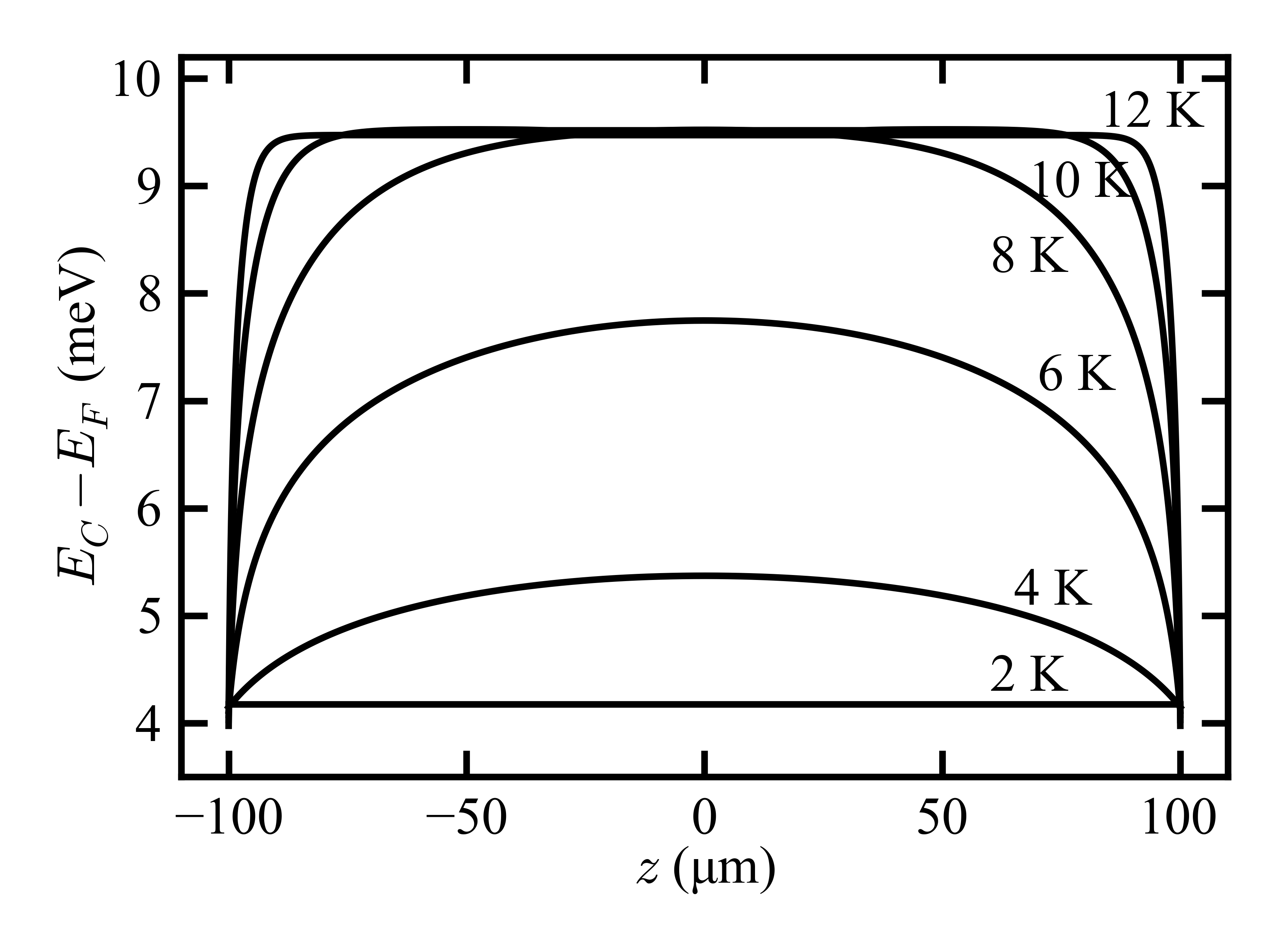}%
\caption{Calculated bulk conduction band at various temperatures for $E_{pin} = 5.5$ meV. \label{fig4}}
\end{figure}

Fig. \ref{fig4} shows how the band structure varies with temperature, again using $E_{pin} = 5.5$ meV; the valence band is omitted. At 12 K, the highest temperature shown, the conduction band for the majority of the bulk is $E_{gap}/2$ above the Fermi energy. There is a small amount of band bending at the edges, but it does not extend very far into the bulk, as expected from Eq. \ref{lengthscale}. This means that the band structure is similar to that of a standard gapped material, except near the surface. As the temperature is lowered, however, the band bending effects begin to extend farther into the bulk. At 2 K, the lowest temperature shown, these effects completely dominate the band structure. Here, the conduction band is much closer to the Fermi energy than the valence band is, and this result is very different from what is observed in a standard gapped material. 

This process demonstrates a crossover between bulk conduction dominated by the usual bulk effects (at high temperatures) and bulk conduction dominated by surface effects (at low temperature). We can understand where the crossover occurs by comparing the charge densities and relevant length scales. At high temperatures, the bulk dominates bulk transport, and this can be characterized by the size of the sample ($t$) and the intrinsic carrier density ($n_i$). At low temperatures, the surface dominates bulk transport, and this can be characterized by the depletion length ($l$) and the carrier density on the surface ($n_s$). From this, we can estimate the crossover as occurring when $t n_i \approx l n_s$. In our calculation, we estimate that the crossover occurs at about 10-12 K. Since the crossover depends on the relationship between the depletion length and the sample, the crossover temperature is dependent on the thickness of the sample. For example, in a much thinner sample such as a thin film, the crossover would occur at a higher temperature than we have estimated in this calculation. 

These band bending results and the crossover are able to explain the gap discrepancy between transport and spectroscopy. In our picture, spectroscopy still measures the full gap, which does not change based on where the measurement occurs in the sample. Transport, on the other hand, measures a different activation gap depending on temperature. Below the crossover, transport measures the activation gap on the surface, as shown in Fig. 3(a), but above the crossover, transport measures the average gap across the entire bulk. As a rough estimate using our model's parameters, at 4 K the total gap is about 19 meV, and with a pinning of 5.5 meV the model yields $E_a = 4$ meV, in agreement with experiment. In this way, the gap discrepancy can be understood without using in-gap bulk impurity states. 

The crossover also has interesting implications for the gap. As we have noted, the depletion length extends through the bulk at low temperatures. As the temperature is lowered, the depletion length diverges, and this can be thought of in analogy with semiconductors. In semiconductors, depletion length increases with purity, and a completely pure material would have an infinite depletion length.\cite{sze} Our calculation shows that the depletion length becomes large at cryogenic temperatures in \smb{}, consistent with the hypothesis that the gap is clean and the bulk is truly insulating. 

We note that the Fermi energy pinning is not rigid, which means that fixing the conduction band permanently at $E_{pin}$ is not exact for all temperatures. Near the crossover, the Fermi energy is allowed to shift slightly. This means that competing with the surface-to-bulk crossover is a slightly shifting pinning. This type of shift does not affect the fully surface or fully bulk regions seen in the model, but if its effects were included, it would slightly change the temperature at which the crossover occurs. We have chosen not to include these effects, because the model is quite robust against changes. The band bending result is always present, and even a large variation in the parameters only slightly shifts the magnitude of the calculated effect. For example, reports of dielectric constant \cite{gorshunov, travaglini} vary from 600 to 1500. When comparing these extremes in the calculation, the results above the crossover is exactly the same and the results below the crossover differ only slightly.

\section{Connection to experiment}
To assess the validity of our model, we must connect the results of the self-consistent calculation to measurable parameters. Specifically, we examine Hall coefficient, resistivity, and thermopower, comparing the simulation results for each to data. Although we have used a semiconductor picture, \smb{} is very different from a standard semiconductor due to its non-parabolic dispersion arising from the hybridized band structure. Because of this unusual band structure, its transport properties are unique and must be considered in detail. We must consider which carriers, $d$-like or $f$-like, contribute to transport phenomena, as well as the sign of these carriers. 

In the discussion of the simplified DOS above, we saw that the $f$-like electrons dominate when calculating the carrier density, because their contribution to the DOS is much greater than the contribution of the $d$-like electrons. However, the $f$-like electrons have a flat dispersion, which must yield zero mobility. This means they cannot contribute to transport, so observed transport phenomena must be due to the $d$-like electrons. Looking back at the features of the actual band structure, we note that the $f$-like electrons do not have an exactly flat dispersion, and there is some curvature connecting the $f$- and $d$-like states. However, based on ARPES data, these features are small compared to the size of the gap.\cite{denlinger} Thus the curvature of these features is small compared to the simplified band structure, so we can again neglect the effects.

We can understand this more quantitatively using semiclassical (Boltzmann) transport. First, we will consider the case of an intrinsic semiconductor in the conduction band to demonstrate the calculation, but the valence band result can be found similarly. We will then discuss modifications for the \smb{} case. Using the elementary solution to the Boltzmann equation, the current due to an electric field applied in the $z$-direction is \cite{ziman}
\begin{equation}
\label{current}
j_z = - \int_{E_C}^{\infty} v_z^2 e^2 \tau E_z \frac{\partial f^0 (\epsilon)}{\partial \epsilon} g_C(\epsilon) d \epsilon
\end{equation}
where $v_z$ is the particle velocity, $e$ is the electronic charge, $\tau$ is the scattering time, $E_z$ is the applied electric field, $g_C(\epsilon)$ denotes the DOS for the conduction band, and $\epsilon$ is the energy. We use the relaxation time approximation, where $\tau$ is independent of energy, and the equipartition theorem, $v_z^2 = v^2/3$, to rewrite Eq. \ref{current}. Then, using $j_z = \sigma E_z$, we find the conductivity, 
\begin{equation}
\label{beforeident}
\sigma \approx - \frac{e^2 \tau}{3} \int_{E_C}^{\infty} v^2 \frac{\partial f^0 (\epsilon)}{\partial \epsilon} g_C(\epsilon) d \epsilon. 
\end{equation}
The derivative of the Fermi-Dirac distribution is \cite{ziman}
\begin{equation}
\label{ident}
\frac{\partial f^0 (\epsilon)}{\partial \epsilon} = - \frac{1}{\kt} f^0 (\epsilon) (1-f^0 (\epsilon))
\end{equation}
and for a general intrinsic semiconductor, the conduction band is almost empty, so the term in parentheses can be approximated as 1. Then the conductivity becomes
\begin{equation}
\label{withv}
\sigma \approx \frac{e^2 \tau}{3\kt} \int_{E_C}^{\infty}  v^2 f^0 (\epsilon) g_C(\epsilon) d \epsilon.
\end{equation}
Since only electrons near the Fermi energy are mobile, we can approximate their velocity as the Fermi velocity $v_F$, and this is a constant. The remaining integral is just the usual method for calculating carrier density, so we find
\begin{equation}
\label{nov}
\sigma \approx \frac{e^2 \tau}{3\kt} \:  v_F^2   n
\end{equation}
where $n$ is given by Eq. \ref{densn}. To further simplify, we can use the Einstein relation for semiconductors, which relates the diffusion constant, $D = v^2 \tau/3$, to the mobility by 
\begin{equation}
\frac{\mu \kt}{e} = \frac{v^2 \tau}{3}
\end{equation}
where $\mu$ is the mobility, $v$ is the average velocity, and 3 represents the number of dimensions (the right-hand side Fof this equation can be derived using the equipartition theorem). So we find, for average velocity $v_F$, the familiar result, written for an intrinsic semiconductor, 
\begin{equation}
\label{sig}
\sigma = n e \mu.
\end{equation}

In \smb{}, the picture is a little more involved. We will now re-derive the general result of Eq. \ref{sig} with modifications for \smb{}. First, we consider the carriers. Since there are two types of carriers ($f$-like and $d$-like), a small displacement of the Fermi surface due to an applied electric field is not uniform. For our simplified dispersion (Fig. 1c), say the field is being applied from right to left (so that electrons move from left to right). Then the electrons in $d$-like states on the right are mobile as they would be in a conventional semiconductor. However, the electrons in $d$-like states on the left are unable to move, as the $f$-like states are filled and have zero mobility. This means that only half of the carriers in the band can move when a current is present. Therefore we must include a factor of $1/2$ relative to the usual result (Eq. \ref{ident}). 

Additionally, since only the $d$-like carriers contribute to transport, we should only consider the carrier density $n$ coming from these. Using Eq. \ref{densn} and Eq. \ref{ncd}, we find that 
\begin{equation}
n =  g_0 \kt e^{-(E_C - E_F)/\kt}.
\end{equation}
Now we can rewrite $g_0$ using Eq. \ref{dos3d}. Also, the carriers should still move with an average speed of $v_F$. We obtain 
\begin{equation}
\sigma_{\scriptsize{\textrm{SmB}}_6} \approx \frac{e^2 \tau v_F^2}{6} \: \frac{m}{\pi^2 \hbar^3} \sqrt{ 2m E_F}\: e^{-(E_C - E_F)/\kt}
\end{equation}
and this can be further simplified using $\sqrt{2m E_F} = \hbar k_F$ and $v_F = \hbar k_F / m$ to obtain
\begin{equation}
\sigma_{\scriptsize{\textrm{SmB}}_6} \approx \frac{e^2 \tau k_F^3}{6 m \pi^2} \: e^{-(E_C - E_F)/\kt}.
\end{equation}
Next, we apply $k_F = (3 \pi^2 n)^{1/3}$, which can be calculated by integrating to find the carrier density at zero temperature and rearranging. This means that $n$ in this expression is the density of filled states up to $E_F$, and according to Eq. \ref{ncf}, this is just $n_{ell}$. We also use $\mu = e \tau /m$ to find 
\begin{equation}
\label{sigmac}
\sigma_{\scriptsize{\textrm{SmB}}_6} \approx \frac{1}{2} n_{ell}  \: e  \: \mu_{d} \: e^{-(E_C - E_F)/\kt}
\end{equation}
where the subscript on mobility denotes that only $d$-like electrons are mobile.

This calculation can be repeated for the valence band, and the result is similar, except that the exponential is replaced by $\exp{[-(E_F - E_V)/\kt]}$. To understand how the conduction and valence band contributions are related physically, we must consider the signs of the carriers in both bands. For a conventional semiconductor, the conduction band contributes electrons with positive effective mass and the valence band contributes holes with negative effective mass. These have opposite contributions to transport. In \smb{}, we still have electrons in the conduction band and holes in the valence band, but both bands have positive curvature in the $d$-like electrons. Since we are just considering the $d$-like carriers, the conduction band case is the same as that of a conventional semiconductor, electrons with positive effective mass. However, the valence band has positive curvature rather than negative as it would for a conventional semiconductor. This means that although there are holes in the valence band, they have a positive effective mass as well, so they contribute with the same sign as the electrons in the conduction band. 

Although this result was found by considering the simplified band structure, again we can neglect the details of the bands. For example, the $H$-point feature in the valence band observed by ARPES \cite{denlinger} shows up as a small bump with negative curvature in the $f$-like part of the dispersion. As discussed previously, this feature is very close to the valence band, so although it creates some curvature in the valence band, the effect is small. This means that there are holes with negative effective mass at these points in the valence band, but we assume that the curvature is large, so that these carriers have a much smaller mobility than the $d$-like carriers.

Because the dominant ($d$-like) carriers in the valence and conduction band contribute to transport with the same sign, we can return to Eq. \ref{sigmac}  and conclude that the total conductivity for all carriers in both bands is the usual result as in Eq. \ref{sig}, with $\mu = \mu_d$, and $n$ defined as
\begin{equation}
\label{carrydens}
n = \frac{1}{2} n_{ell}  \: e^{-(E_C - E_F)/\kt} + \frac{1}{2} n_{ell} \: e^{-(E_F - E_V)/\kt}
\end{equation}
This means that we can use the usual transport relations to connect our model to the experimental results, provided that this expression is used to calculate the carrier density.

\section{Transport in the model}

\subsection{Hall coeffcient and resistivity}

\begin{figure}
\includegraphics[scale=1, trim = 2mm 4mm 0mm 0mm, clip]{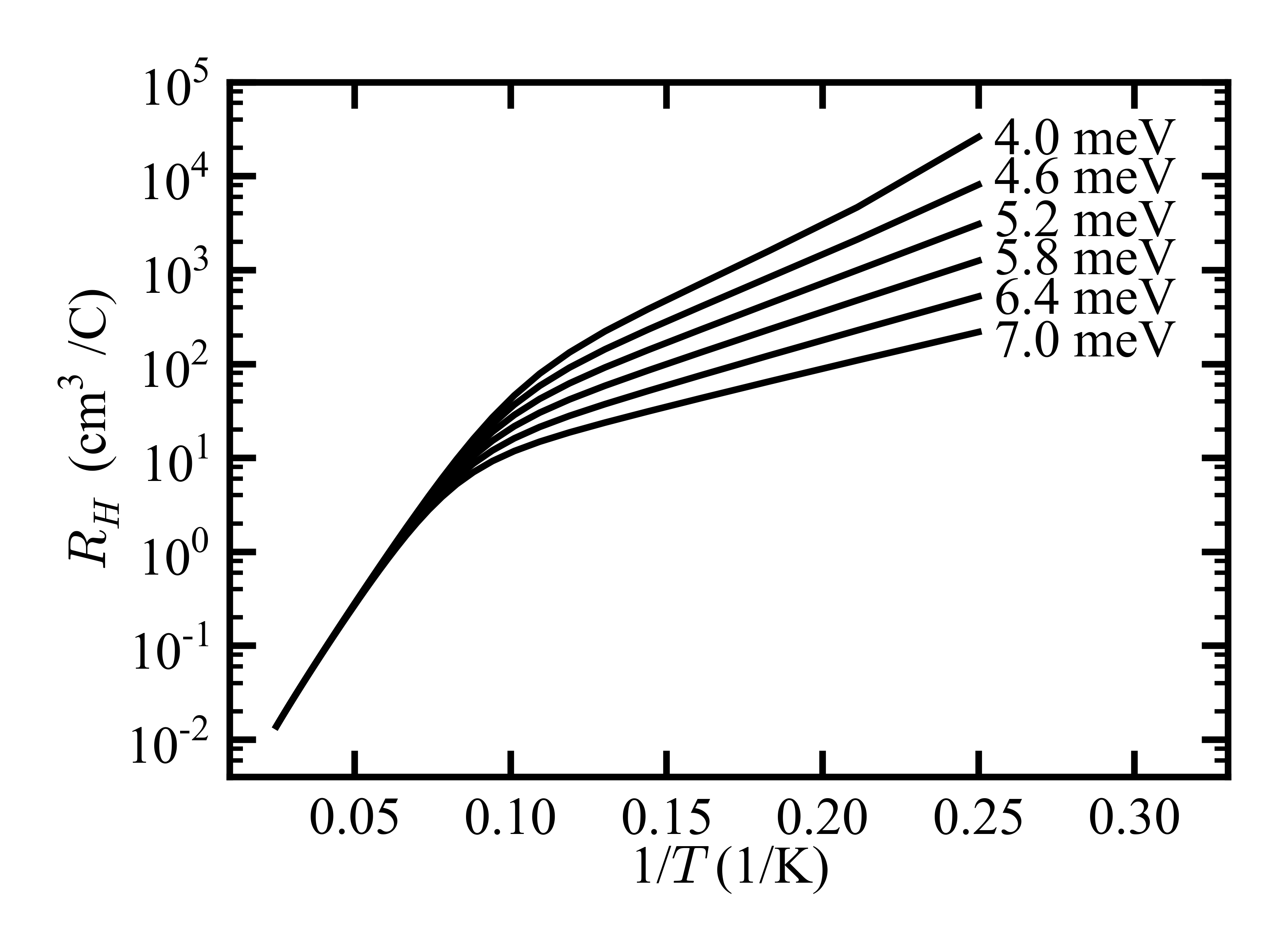}%
\caption{Calculated Hall coefficient as a function of temperature for different values of $E_{pin}$.   \label{fig5}}
\end{figure}

We can now combine the results $E_C (z)$  and $E_V (z)$ of the self-consistent calculation with Eq. \ref{carrydens} to define an effective carrier density
\begin{equation}
\begin{split}
n_{eff} = \frac{1}{t}  \int_{-t/2}^{t/2} & \: \:\frac{1}{2} n_{ell} \:  \bigg( e^{-[E_C(z) - E_F]/\kt} \\
&+ \: e^{-[E_F - E_V(z)]/\kt} \bigg) dz
\label{neff}
\end{split}
\end{equation}
where $t$ is the thickness of the sample ($t=200 \: \mu$m for our test sample). We can then use this carrier density to compare the model to transport data. We first concentrate on the Hall coefficient ($R_H=1/ne$) because it does not require any further parameters to be included; however, if we assume that mobility is constant, the resistivity follows the same trend. This is not a good assumption, as mobility is often temperature dependent, but the same feature around 10 K is seen in data for both Hall coefficient and resistivity.

Fig. \ref{fig5} shows a plot of calculated Hall coefficient as a function of temperature for various values of $E_{pin}$. As in the band structure result of Fig. \ref{fig4}, we observe a crossover around 10 K. At temperatures above this crossover, where the bulk transport is dominated by bulk effects, all values of $E_{pin}$ yield the same curve. This is expected because in this region, the depletion length is always much less than the sample size, regardless of $E_{pin}$. Below the crossover, however, there is some variation. In this region, the amount of bending influences the depletion length, so the magnitude of the Hall coefficient changes with $E_{pin}$. As mentioned previously, in this region, the activation energy can also be determined by $E_a = E_{gap}/2 - E_{pin}$.

\begin{figure}
\includegraphics[scale=1, trim = 2mm 4mm 0mm 0mm, clip]{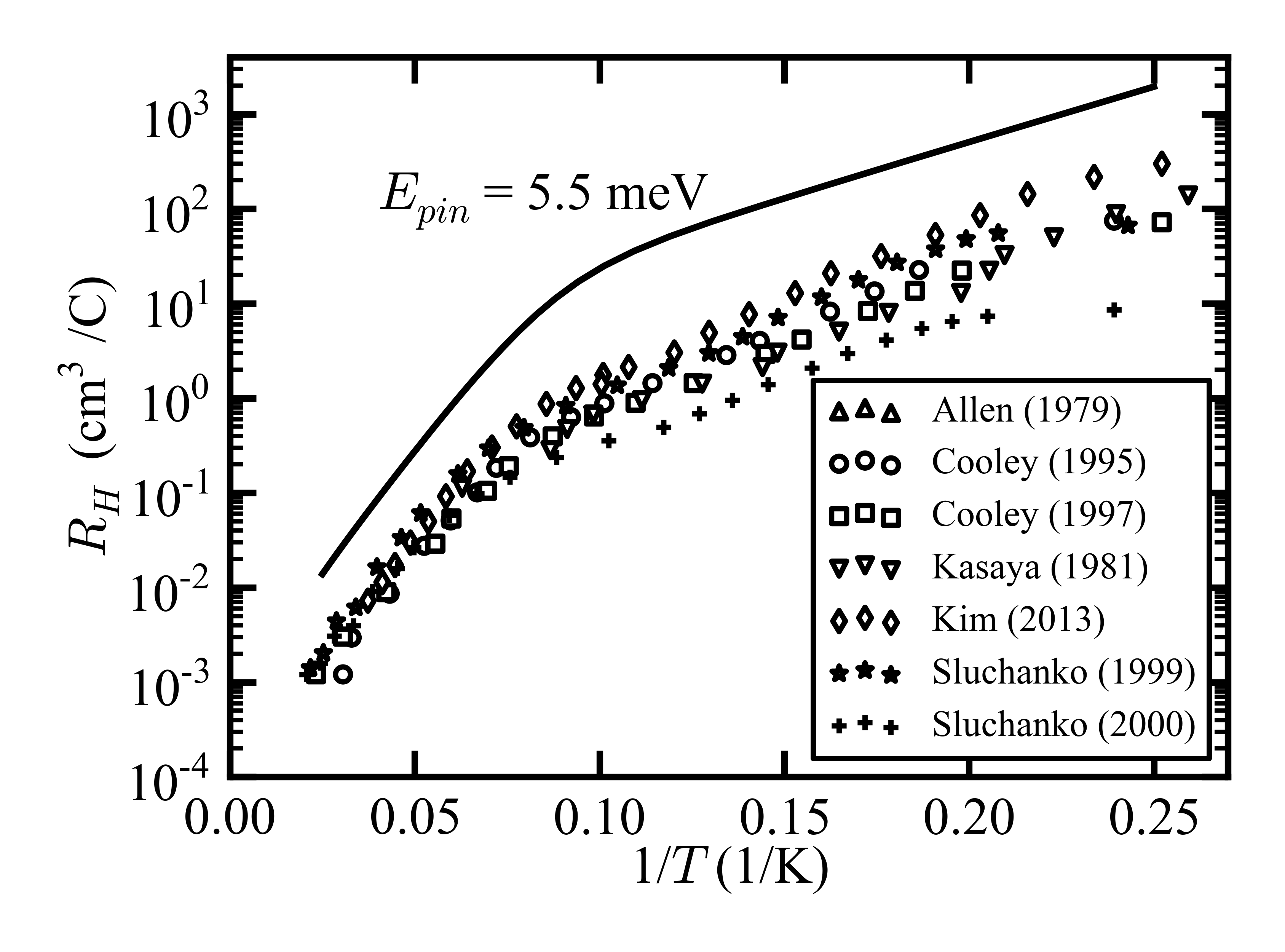}%
\caption{Comparison of one value of $E_{pin}$ to a collection of Hall data. \cite{allen, cooley, cooleythesis, kasaya, kim, sluchanko99, sluchanko00}  \label{fig6}}
\end{figure}

The calculated Hall coefficient for $E_{pin} = 5.5$ meV  is plotted along a collection of data in Fig. \ref{fig6}. The data agrees very well above about 10-12 K, and this agreement suggests that the bulk carrier density in \smb{} is fairly consistent across samples. Below about 10-12 K, the data exhibits a variation of about an order of magnitude. We do not suggest a mechanism for this variation, but we do note that in this temperature range, our model suggests that the bulk conduction is dominated by the surface. Around 10-12 K, a feature is observed by all researchers, and this corresponds to the crossover discussed previously, between bulk conduction dominated by bulk effects above the crossover and bulk conduction dominated by surface effects below the crossover. 

The calculated Hall coefficient from our model also demonstrates a crossover at about 10-12 K and correctly describes the shape of the data, although the magnitude of our result does not agree with the data. This disagreement was expected based on the simplifications made to the dispersion. Previously, we estimated that these simplifications yielded errors of not more than a few meV in the dispersion, but this becomes very important for the Hall coefficient calculation. We can estimate the size of the expected discrepancy in the calculated Hall coefficient by using a Boltzmann factor, $e^{\Gamma/\kt}$, where $\Gamma$ is the approximate width of the features that were neglected. For $\Gamma$ in the range of 1 to 3 meV in a temperature range of 10 to 20 K, the model is expected to be off at least by a factor of 2 and at most by a factor of 32. Our result was consistently a factor of about 5-6 greater than the data, and this is within the expected range of the discrepancy. Based on this estimate, agreement would likely be improved by including more features of the dispersion and DOS. However, this type of refinement would require many more parameters to be introduced. 

\subsection{Thermopower}
We also compare the model to thermopower data, because like the Hall coefficient, it does not require any further parameters. In the relaxation time approximation, thermopower for electrons in a semiconductor is given by \cite{ziman}
\begin{equation}
S_C = -\frac{k_B}{e} \bigg[ \bigg(\alpha + \frac{5}{2}\bigg) - \frac{E_C - E_F}{\kt} \bigg]
\end{equation}
where the subscript denotes the conduction band, and $\alpha$ is a constant between 0 and 2 that describes how energy is related to scattering time ($\tau \propto E^{\alpha}$). A similar expression can be obtained for holes in the valence band; it is important to note that the sign is positive for holes in a standard semiconductor. For a material containing both electrons and holes, these can be combined according to
\begin{equation}
S_{tot} = \frac{S_C \sigma_C + S_V \sigma_V}{\sigma_C + \sigma_V}.
\label{mott}
\end{equation}
In the limit of an intrinsic semiconductor, where $n=p$, the intrinsic carrier density in Eq. \ref{intrinsic} and the usual conductivity in Eq. \ref{sig} can be used to simplify this. Assuming a quadratic dispersion for both the valence and conduction bands, we obtain
\begin{equation}
S = \frac{k_B}{e} \bigg[ \frac{b-1}{b+1} \: \frac{E_{gap}}{2 \kt} + \frac{3}{4} \: \ln{\frac{m_n}{m_p}} \bigg]
\end{equation}
where $b = \mu_n / \mu_p$, and the subscripts refer to electrons ($n$) and holes ($p$).  \cite{ziman, sluchanko99}

\begin{figure}
\includegraphics[scale=1, trim = 2mm 4mm 0mm 0mm, clip]{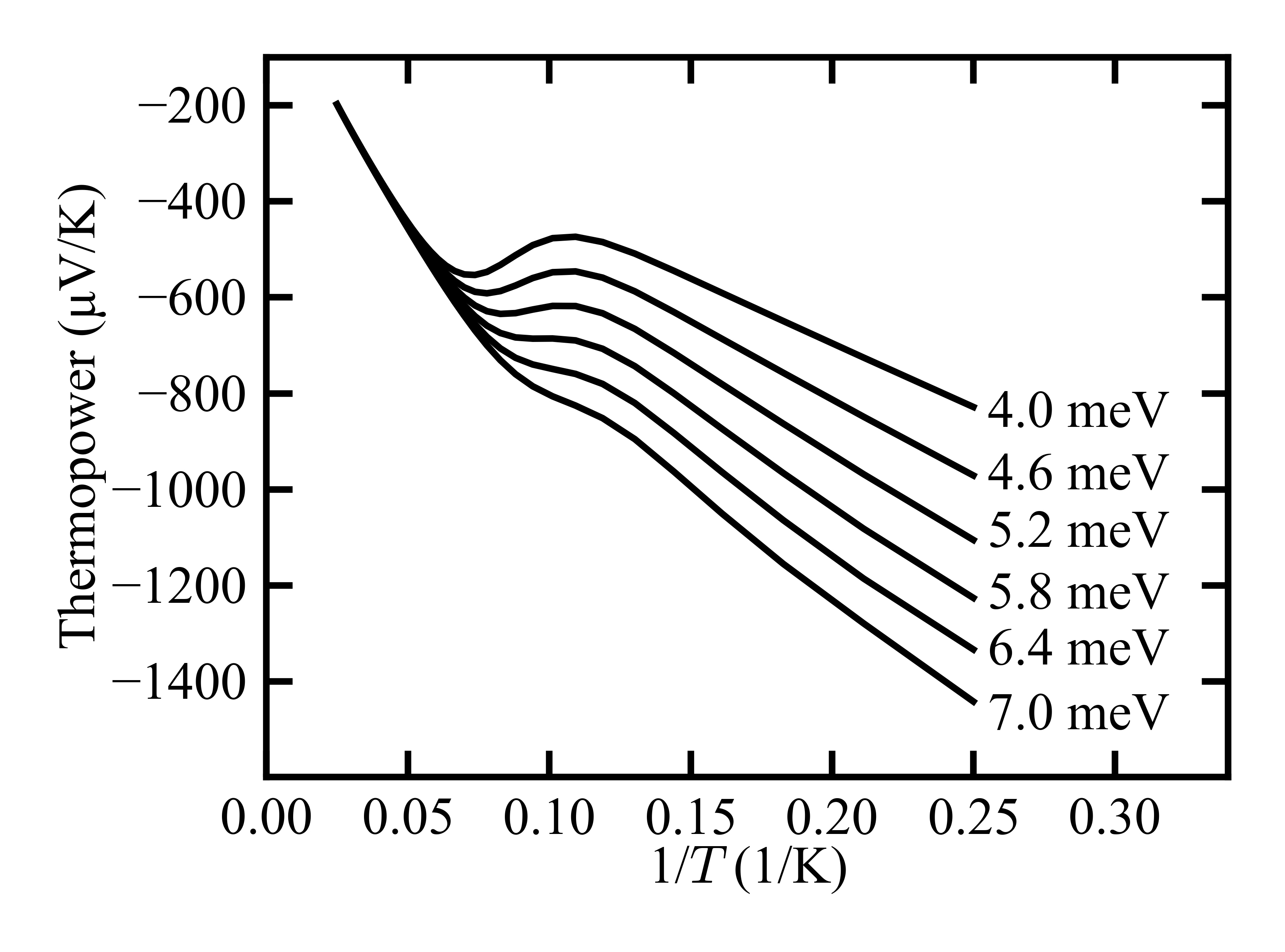}%
\caption{Thermopower versus temperature for various values of $E_{pin}$. A feature can be seen around 10 K, as in the Hall plot. \label{fig7}}
\end{figure}

Again, the picture is slightly different in \smb{}. The factor of 1/2 must be accounted for once again, but based on the form of Eq. \ref{mott}, it is clear that this factor cancels out for thermopower. Also, the curvature of the valence band means that the holes contribute with the same sign as the electrons. Using the conductivity derived in Eq. \ref{sigmac} and the analogous result for the valence band in Eq. \ref{mott}, we can find the total thermopower in the model. Defining $\epsilon_c (z) = E_C (z) - E_F$ and $\epsilon_v (z) = E_F - E_V (z)$, we obtain
\begin{equation}
S(z) = - \frac{k_B}{e} \frac{1}{\kt} \frac{\epsilon_c (z) e^{-\epsilon_c (z)/\kt} + \epsilon_v (z) e^{\epsilon_v (z)/\kt}}{e^{\epsilon_c (z)/\kt} + e^{\epsilon_v (z)/\kt}}
\end{equation}
where $z$ again refers to the location in the bulk in the model. To get the total thermopower, we must integrate this expression, but since thermopower is dependent on conductivity and conductivity is dependent on $z$, it must be integrated using a form similar to that of Eq. \ref{mott}. Thus, the effective thermopower across the bulk is
\begin{equation}
S_{eff} = \frac{\int_{-t/2}^{t/2} S(z) \sigma(z) dz} {\int_{-t/2}^{t/2} \sigma(z) dz}
\end{equation}
where $t$ is the thickness of the sample (here $t=200 \: \mu$m) and $\sigma(z) = n(z) e \mu$, where $n(z)$ is given by the integrand of Eq. \ref{neff}.

$S_{eff}$ was calculated as a function of temperature for various values of the pinning, shown in Fig. \ref{fig7}. Again, a feature around about 10 K is evident, although it is broader than the feature seen for the Hall effect. The bulk behavior in Fig. \ref{fig7} is the same for all values of $E_{pin}$, but the surface behavior and prominence of the feature varies. 

\begin{figure}
\includegraphics[scale=1, trim = 2mm 4mm 0mm 0mm, clip]{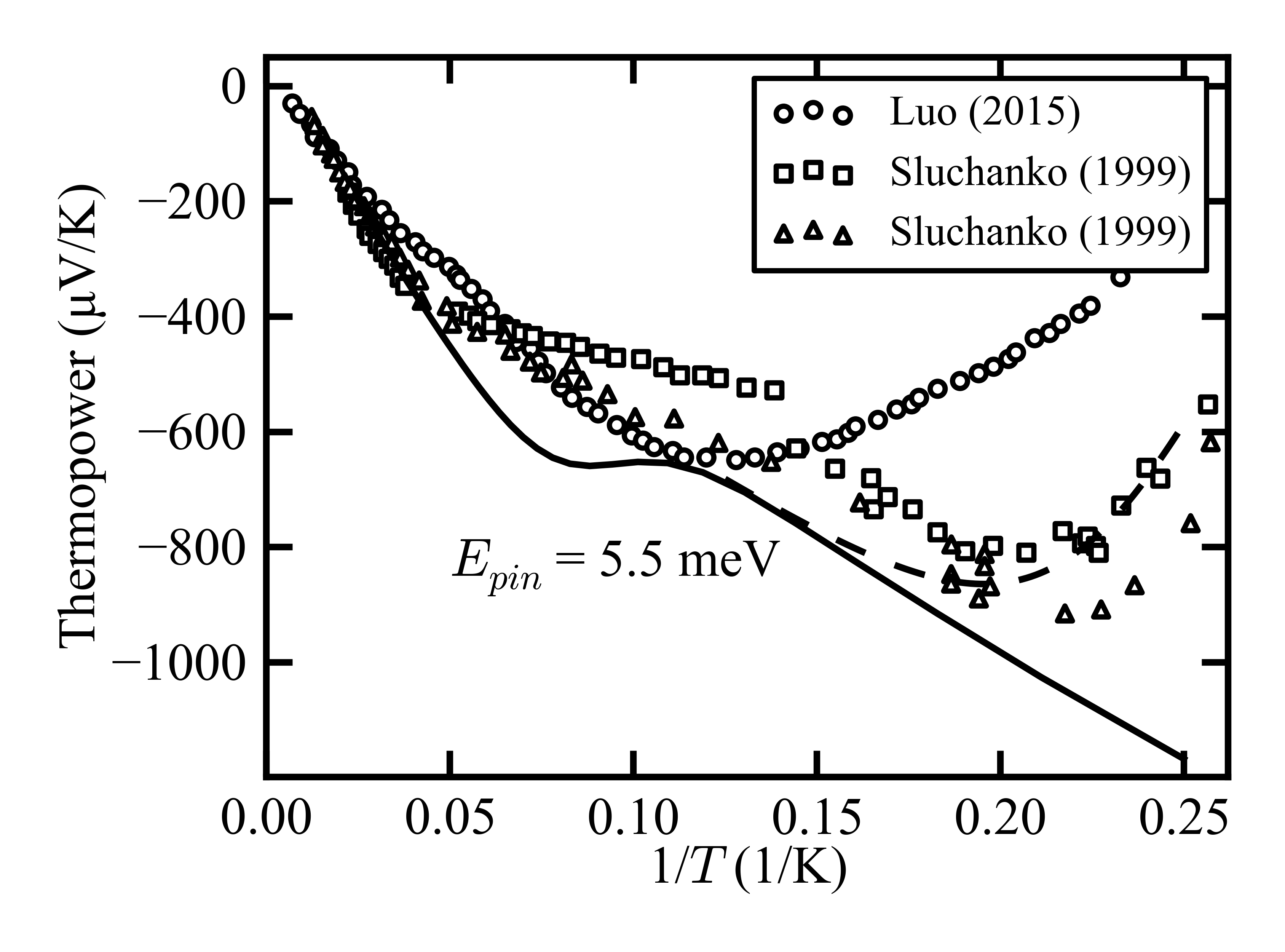}%
\caption{Comparison of calculation for one value of $E_{pin}$ (solid line) to a collection of thermopower data. \cite{luo, sluchanko99} The effects of the TI surface state crossover are also shown (dashed line). \label{fig8}} 
\end{figure}

Fig. \ref{fig8} shows a collection of thermopower data with a fit from the model. The data are consistent at high temperatures, and in this regime, there is also excellent agreement between the data and the model. As with the Hall coefficient, this agreement is expected as bulk effects dominate in this regime. At low temperatures, the data are more diverse. We note that the crossover does not occur at the same temperature in each data set shown, but we do not propose a mechanism for this.

At high temperature, in the bulk-dominated regime, there is good agreement between the data and the model. A feature is also present in both the data and the model, although it does not occur at exactly the same temperature for each set of data. At low temperature, a deviation can be observed in the data, although it does not occur at the same point in each data set. We attribute this deviation to the manifestation of the TI surface state near 4 K, which can be added to the calculation. 

To estimate the contribution of the surface state, we will again use Eq. \ref{mott},
\begin{equation}
S_{tot} = \frac{S_b \sigma_b t + S_s \sigma_s}{\sigma_b t + \sigma_s},
\end{equation}
where the subscripts $b$ and $s$ refer to bulk and surface contributions, respectively, and the thickness $t$ is included so that the units match. From the theoretical treatment of TI surface states, \cite{takahashi} we expect that their contribution is much smaller than the bulk contribution. Therefore, only the first term will have a significant contribution to the thermopower. Near the bulk-to-surface transport crossover of $T_c = 4$ K, we also expect that the bulk and surface will contribute similarly to transport, i.e. $\sigma_b t \approx \sigma_s$. Each of these can be approximated using the form $\sigma_0 e^{E_a/\kt}$, where $E_a$ is the activation energy (the energy relevant to transport). For $\sigma_s$ we use $T_c =4$ K in this expression, and for $\sigma_b t$ we allow $T$ to vary. We also assume that near $T_c$, $\sigma_0$ is about the same for both surface and bulk contributions.  Then the thermopower near 4 K, with $E_a = 3.47$ meV, \cite{wolgast} is approximately
\begin{equation}
S_{tot} \approx \frac{ e^{E_a/\kt}}{ e^{E_a/\kt}+ e^{E_a/k_B T_c}} \:  S_b.
\label{surf}
\end{equation}
This expression is shown in Fig. \ref{fig8} with a dashed line, and the result provides a better estimate of the data near 4 K than the fit from the calculation.

Our model, with the addition of the estimation of surface effects, captures the low-temperature features of the thermopower well. We note that there is not much thermopower data available in the literature, which makes it difficult to understand trends in the data as done for the Hall coefficient. Our model agrees quite well with the data from Sluchanko $et al.$ \cite{sluchanko99}, but not as well with the other data. Again, a discrepancy between the data and the model is present, and again we attribute this to neglecting the small features of the dispersion. Improved agreement could likely be attained by adding more details of the dispersion to the model.

\section{Conclusion}
We have presented a new model to understand the difference between the spectroscopy and transport gaps in \smb{} without relying on in-gap bulk states. Transport measures an activation energy of 3-4 meV, or a gap of 6-8 meV, while spectroscopic methods measure a gap of 16-20 meV. This discrepancy between the two results has often been explained by introducing a localized bulk state in the gap, perhaps due to impurities, although other explanations such as the presence of an indirect gap have also been used. The effective mass approximation has been used to understand such an in-gap impurity state, because this method has been successful at describing impurity states in many gapped materials. However, we showed that the effective mass approximation fails when it is applied to \smb{}, suggesting that the in-gap impurity state picture is not justified. 

Instead, we suggested a new way of understanding the \smb{} gap using self-consistent band-bending calculations. We simplified the well-known dispersion and corresponding density of states to capture the main characteristics of \smb{} and modeled \smb{} as an intrinsic semiconductor. We considered the possibility of band bending, which is expected to arise from the presence of excess surface charges. A self-consistent solution for the potential was found by numerically solving Poisson's equation with this charge density across the bulk of a test sample, with the boundary conditions simulating the strength of the bending effects at the surface and enforcing symmetry across the bulk. 

The self-consistent solution was found for temperatures from 4-40 K, and from this result the band structure was calculated. In addition, the band structure result was connected to measurable transport parameters using semiclassical transport theory modified for \smb{}. Specifically, we concentrated on Hall coefficient, resistivity, and thermopower, as these do not require the addition of adjustable parameters to our model. The results of these calculations demonstrated a crossover to bulk transport dominated by bulk effects (in analogy with a standard gapped material) at high temperatures, to bulk transport dominated by surface effects at low temperature. The calculated crossover temperature was 10-12 K, which accounts for a feature that has been observed in transport data near this temperature. The transport parameters were compared to data, and although there was qualitative agreement between the data and the calculated curves, their magnitudes did not agree. We attributed the disagreement to neglecting some of the small features of the actual dispersion in the model, and found that a few meV of error in the dispersion could explain the disagreement.

We also related the feature seen in transport data to the depletion length for a semiconductor. At high temperatures, the depletion length was much smaller than the sample size, and at low temperatures, the depletion length was much larger than the sample size. We estimated that the crossover would occur when bulk effects in terms of sample thickness and intrinsic carrier density become comparable to surface effects in terms of depletion length and surface charges.  Because of this relationship between depletion length and sample size, our model suggests that the crossover would occur at different temperatures for different sample thicknesses. For example, we predict that for an \smb{} thin film, the crossover temperature would be higher than the 10-12 K temperature we calculated. Our model could be further tested by examining the thickness dependence of the crossover (i.e. the feature around 10 K observed in transport). 

Additionally, the divergence of the depletion length in \smb{} at low temperatures suggests that the gap is clean, similar to the gap in a superconductor. This, combined with the success of our model at describing a variety of data without introducing bulk states in the gap, agrees well with the observation that there is no residual bulk conduction in \smb{} below the TI crossover temperature of about 4 K. Together, these observations imply that \smb{} is a true TI; it does not exhibit bulk conduction below the TI crossover temperature as all other known TIs do. This would be exciting for research in technological applications that require a clean gap and no bulk conduction. We also predict that our model could be extended to other materials that have a dispersion similar to that of \smb{}, including alloys of \smb{}.

\begin{acknowledgments}
The authors thank Lu Li and Steven Wolgast for helpful discussions, as well as J.W. Allen and Jonathan Denlinger for helpful discussions and advice on improving the manuscript. Funding for this work was provided by NSF grants \#DMR-1441965 and \#DMR-1643145.
\end{acknowledgments}

\bibliography{bendbib}

\end{document}